\documentstyle[12pt,aasms4]{article}

\input epsf.sty

\newcommand{\etal}{et~al.\ }
\newcommand{\eg}{e.g.\ }
\newcommand{\ie}{i.e.\ }
\newcommand{\Msun}{M_{\odot}}

\newcommand{\kms}{km~s$^{-1}$}

\newcommand{\HeI}{He~{\sc i}}
\newcommand{\OI}{O~{\sc i}}
\newcommand{\OII}{O~{\sc ii}}

\newcommand{\NaI}{Na~{\sc i}}

\newcommand{\MgI}{Mg~{\sc i}}

\newcommand{\CaII}{Ca~{\sc ii}}

\newcommand{\FeI}{Fe~{\sc i}}
\newcommand{\FeII}{Fe~{\sc ii}}
\newcommand{\FeIII}{Fe~{\sc iii}}

\newcommand{\CoIII}{Co~{\sc iii}}

\newcommand{\Fefs}{$^{56}$Fe}
\newcommand{\Cofs}{$^{56}$Co}
\newcommand{\Nifs}{$^{56}$Ni}
\newcommand{\Mej}{$M_{\rm ej}$}
\newcommand{\KE}{$E_{\rm kin}$}

\begin{document}

\lefthead{Mazzali}
\righthead{Asymmetry in SN~1998bw}

\title{The nebular spectra \\
	of the hypernova SN 1998bw \\
	and evidence for asymmetry \\ }

\author{Paolo A. Mazzali\altaffilmark{1,2}, 
	Ken'ichi Nomoto\altaffilmark{3,2}, \\
	Ferdinando Patat\altaffilmark{4}, and 
	Keiichi Maeda\altaffilmark{3} \\ } 
	
\altaffiltext{1}{Osservatorio Astronomico di Trieste, 
  via G. B. Tiepolo 11, I-34131 Trieste, Italy; mazzali@ts.astro.it}
\altaffiltext{2}{Research Center for the Early Universe,
  School of Science, University of Tokyo, Tokyo 113-0033, Japan}
\altaffiltext{3}{Department of Astronomy, School of Science, 
  University of Tokyo, Tokyo 113-0033, Japan; nomoto@astron.s.u-tokyo.ac.jp}
\altaffiltext{4}{European Southern Observatory, 
 Karl-Schwarzschildstr. 2, D-85748 Garching, Germany; fpatat@eso.org}

\begin{abstract}
The nebular spectra of the energetic Type Ic supernova SN~1998bw (hypernova) 
are studied. The transition to the nebular phase occurred at an epoch of about
100 days after outburst, which is assumed to coincide with GRB980425. 
Early in the nebular epoch the spectra show the characteristics of a typical 
SN~Ic spectrum, with strong lines of [\OI], \CaII\ and \MgI], and lines
of [\FeII].  However, the [\FeII] lines are unusually strong for a SN~Ic. 
Also, lines of different elements have different widths, indicating different
expansion velocities. In particular, iron appears to expand more rapidly than 
oxygen.
Furthermore, the [\OI] nebular lines decline more slowly than the [\FeII] ones,
signalling deposition of $\gamma$-rays in a slowly-moving O-dominated region. 
These facts suggest that the explosion was aspherical.  The absence of 
[\FeIII] nebular lines can be understood if the ejecta are significantly 
clumped. A schematic picture of what this very unusual stellar explosion 
may have looked like is presented.
\end{abstract}


\section{Introduction} 
 
SN~1998bw was a very interesting object from the word go (\eg Nomoto
\etal 2001). Discovered in the
error box of GRB980425, and very possibly linked to it, this object was soon
identified to be a supernova (SN) from its light curve, which was different
from those of typical optical transients of GRB's. Extensive data were 
collected at ESO, where SN~1998bw was designated a Target of Opportunity 
(Galama \etal 1998; Patat \etal 2001), and at other observatories. 

Early spectra were rather blue and featureless, quite unlike those of other 
known SNe. However, a somewhat more careful inspection showed clear 
similarities with the spectra of Type Ic SNe (SNe~Ic), but with one major 
difference: the absorption lines were so broad in SN~1998bw that they blended 
together, giving rise to what could even be confused with an emission 
spectrum. Very large expansion velocities ($\sim 30,000$~km~s$^{-1}$) were 
measured in the ejecta. Also, SN~1998bw was very bright for a SN~Ic: for a 
redshift distance of 39~Mpc ($z=0.0085$ and $H_0 = 65$~km~s$^{-1}$~Mpc$^{-1}$), 
the SN reached a maximum V$= -19.4$ mag, which is bright even for a SN~Ia, 
while a typical SN~Ic like SN~1994\,I was almost 2 mag dimmer (\eg
Nomoto \etal 1994; Richmond \etal 1996).

Modelling of the early data (Iwamoto \etal 1998, hereafter IMN98) led to 
several striking conclusions. First, the SN produced about $0.7 \Msun$\ of
\Nifs, as much as a SN~Ia. This was necessary to power the light curve via the
deposition of the $\gamma$-rays emitted in the decay of \Nifs\ into \Cofs\ and
\Fefs.

IMN98 discuss that the light curves are degenerate, \ie different combinations
of ejecta mass \Mej\ and kinetic energy \KE\ can all reproduce the data equally
well. Spectra came to the rescue then: spectrum synthesis showed that the only
way to get the observed extensive line blending was to have significant amounts
of material moving at very large velocities. Since models with different \KE\ 
give different degrees of blending, it was possible to select as best model one
with \Mej~$=10.9 \Msun$\ and \KE~$= 3\cdot 10^{52}$~erg. This had an exploding 
core mass of $13.8 \Msun$ and was designated CO138 accordingly. A typical SN~Ic 
has \Mej~$=1 \Msun$, and the typical kinetic energy of most SNe of all types is
only about $10^{51}$~erg (aka 1 foe). An even higher value of 50 foe was
obtained for SN~1998bw by Nakamura \etal (2001a). This gives a better fit to the
declining part of the light curve and to the spectra. A comparable value was 
obtained by Branch (2001) from an analysis of the early time spectra. Because 
of its exceptionally large \KE, SN~1998bw was called a `hypernova'. The 
progenitor must have been a massive star, and we estimated a main sequence mass 
of about $40 \Msun$.  Also, the remnant mass, which is computed by allowing 
only as much of the \Nifs\ produced by core Si burning to be ejected as is 
necessary to power the SN light curve, turned out to be $\sim 3 \Msun$
Nakamura \etal (2001a,b). 
This exceeds the maximum mass of a Neutron Star, suggesting that the explosion 
that was observed as SN~1998bw resulted in the creation of a Black Hole.

Since SN~1998bw was probably connected to a highly non-spherical event like a  
GRB, departure from spherical symmetry could be expected. Early polarization
measurements confirmed this. Polarization of $\sim 0.5$\%, possibly decreasing 
with time, was detected (Kay \etal 1998, IMN98, Patat \etal 2001). This was
interpreted as an axis ratio of about 2:1 in the expanding ejecta (H\"{o}flich
et al. 1999). Models of the collapse of a rotating stripped core of a massive
star were developed, which confirmed that the explosion should be asymmetric,
thus establishing a possible link with the GRB (MacFadyen \& Woosley 1999).

All the above conclusions were based on only about two months worth of data.
Patat \etal (2001) presented extensive data covering more than one year.  
Note that here and throughout the paper epochs refer to the SN time since 
outburst, which is taken to have coincided with GRB980425, corrected for 
time dilation (the host galaxy of SN~1998bw, ESO 184-G82, has a redshift 
$cz = 2532$\,\kms), while Patat \etal (2001) used the time from $B$ 
maximum, which occurred on 10 May, 1998).  
The evolution of SN~1998bw after the first two months reserved several 
surprises: after day 60 the light curve began declining less steeply than 
the model prediction (Mc Kenzie \& Schaefer 1999); the nebular spectrum, 
which started to develop after about day 100, initially appeared to be a 
composite of a SN~Ia and a SN~Ib/c spectrum, showing strong Fe lines in the 
blue, typical of SNe~Ia, and strong O and Ca lines in the red, typical of 
SNe~Ib/c. What was even more peculiar was that there were two families of 
lines: one with a large velocity comprising mostly the \FeII\ lines, and 
another with a smaller velocity ([\OI], \MgI]). 
Also, the various lines declined with different rates, which apparently 
contradicts the hypothesis that the only heating source is the decay of 
\Cofs\ (Patat \etal 2001).

Several papers have addressed the behaviour of the light curve over the first
200 days (Mc Kenzie \& Schaefer 1999), 600 days (Sollerman \etal 2000) and up
to 800 days (Nakamura \etal 2001a). In all these papers the suggestion has been
made that either the density distribution computed in spherically symmetric
explosion models is not correct or that some degree of asymmetry in the 
explosion is probably required to explain the unusual behaviour of the light
curve. In this paper we focus on the late time spectra, and show that they too
require significant asymmetry in the ejecta.

\section{Nebular spectra} 

As discussed by Patat \etal (2001), the transition from photospheric to nebular
spectra took place between about day 65 and day 115, although net emission
may have been present in the \CaII~IR triplet as early as day 37.  
Later other \CaII\ lines, the H\&K doublet and the forbidden 7320\AA\ line 
became visible, followed by lines of [\OI], \MgI], by \NaI~D and by an 
emission at $\sim 5200$\AA. This is most likely due to a complex of [\FeII] 
lines, although at the earliest phases a possible contribution is present 
from permitted \FeII\ transitions, which can produce nebular emission at 
sufficiently high density (Filippenko 1989), and most likely also from
[\OI] 5577\AA.
At about day 115 the spectrum of SN~1998bw was an interesting hybrid 
between that of a SN~Ia (strong [\FeII]) and that of a SN~Ib/c (very strong 
[\OI], \MgI], \CaII). This is already an indication that both Fe and the 
$\alpha$-elements are present with significant abundances in the ejecta. 
Another SN~Ic for which an intense blue flux has been observed
in the nebular epoch is SN~1999dq (Matheson \etal 2000a). This blue
flux has been attributed to \FeII\ lines, but the wavelength
coincidence of this feature with the [\FeII] one in SN~1998bw makes
the [\FeII] identification also a serious possibility. SN~1999cq was a
bright SN~Ic (at least using a redshift distance), so the presence of
a larger amount of Fe than in `normal' SNe~Ic is to be expected. 

The nebular spectra span epochs from 108 to 388 days. 
In all the spectra six dominant features can be identified: \MgI]
at 4500\AA\ (1), the [\FeII] complex near 5250\AA\ (2), \NaI~D at 5900\AA\ (3),
[\OI] 6300\AA\ (4), a [\CaII]-[\OII] blend near 7200\AA\ (5) and the \CaII\ IR
triplet near 8500\AA\ (6).  However, although these features persist for almost
one year, the change in the properties of the spectrum are striking. 
In particular, the [\OI] and \MgI] lines become stronger relative to the 
[\FeII] lines as time goes on, but at the same time they also become narrower.
While in the first spectrum all lines are broad ($v \sim 12000$\,\kms), already 
in the second spectrum [\OI], \MgI] and partly also \CaII\ IR are narrower than 
the [\FeII] complex. This behaviour continues for the duration of our spectral
coverage, throughout which the [\OI] and \MgI] lines and the [\CaII]-[\OII]  
blend grow in strength relative to the other three features. In practice, the 
decline rates of the various emission lines are different, so that while in
the first spectrum features (1), (2), (4) and (6) all have similar peak fluxes,
and feature (5) is as strong as feature (3), in later spectra [\OI] is the
strongest line by far, while the [\FeII] feature rapidly decreases in strength
relative to the other lines, so much so that after about one year the
spectrum of SN~1998bw would be very difficult to distinguish from that of a
normal SN~Ib/c (see Patat \etal 2001 for a direct comparison and
Matheson \etal 2001 for lots of late-time spectra of SNe~Ic). 
At late epochs the strength of the [\OI] and
\MgI] lines is mostly due to the narrow component, which develops with time but
is absent in the [\FeII] feature. All the lines that decline slowly become
eventually significantly narrower than the [\FeII] features. 
The weak \NaI~D line also has a narrow profile, and declines relatively 
slowly, while the \CaII\ IR triplet is almost as broad as the [\FeII] 
feature and declines almost as fast. The decline of the ratio \CaII\
IR/[\CaII]~7300\AA\ is to be expected in a nebula of progressively
decreasing density (Ferland \& Persson 1989). 

The features described above persist for the entire period sampled, the 
only noticeable change being that the [\FeII] feature moves bluewards by
about 130\AA\ between day 108 and day 214. This may be due to a
decreasing contribution of an underlying 'photospheric' component,
which is probably still significant at blue wavelengths in the earlier 
epochs, and to a progressively decreasing contribution of permitted 
\FeII\ emission, as the density decreases, and of [\OI] 5577\AA,
which has a rather high excitation potential (its lower level is the
uppper level of [\OI] 6300\AA), and whose strength should decline as
the electron temperature drops. After day 214, though, the emission
feature stops moving, and its identification as the same [\FeII]
feature observed in SNe~Ia is beyond doubt (Patat \etal 2001, Fig. 11). 
However, the decline rate of the luminosity of the [\FeII] feature remains 
larger than that of both [\OI] and \MgI] at all epochs (Patat \etal 2001, 
Fig. 14), and so this is not just due to changing contributions to the 
[\FeII] feature. 

Another feature which receives different contributions is \NaI~D. 
Early in the nebular phase the ground state [\CoIII] multiplet near
5900\AA\ contributes to this line, as it does in SNe Ia. However, the 
contribution is smaller in SN~1998bw because of the lower degree of 
ionisation. In fact, in our models \NaI~D is responsible for about 80\%
of the observed feature. At later times the contribution of \CoIII\ is
further reduced because \Cofs\ decays to \Fefs. This feature may also
receive a contribution from \HeI\ 5876\AA, which may be excited
non-thermally since He and \Nifs\ coexist in the ejecta (Maeda \etal
2001, Fig.1). 

At all epochs, the strength of the [\FeII] feature is unusually large
for a SN~Ib/c. 
Patat \etal 2001 (Fig.11) compared the nebular spectrum of SN~1998bw with 
that of the SN~Ic 1996aq at an epoch of about 9 months, showing that in 
SN~1998bw the entire blue part of the spectrum, where the [\FeII] lines are 
important, is much stronger relative to the [\OI] and \CaII\ lines in the red. 
Spectra of SNe~Ic at an epoch of about 5 months, corresponding to the epoch 
when the spectrum of SN~1998bw had just turned nebular, have been published 
by A.~Filippenko and collaborators (Filippenko \etal 1990 (SNe 1985F, 1987M); 
1995 (SN~1994I); Matheson \etal 2001 (SNe 1990U, 1990aa)), and a comparison 
by eye of their spectra to those of SN~1998bw confirms the relative 
strength of the [\FeII] feature in SN~1998bw even at the earlier epoch. 

These properties of the late-time spectrum of SN~1998bw are odd in 
many respects. First, all (spherically symmetric) evolution-collapse-explosion
models predict that O, which is found in the unburned part of the ejecta, is
located above Fe in mass coordinates. Therefore O should have a larger velocity
than Fe if anything, but not a smaller one. Secondly, if the spectrum is
powered by $\gamma$-ray deposition, the region of the ejecta which should feel
the heating the most should coexist with Fe, which is mostly the product of the
decay of \Cofs. Therefore, all lines should have a similar width, that of the
Fe lines, and they should decline at about the same rate. This is the case for
SNe~Ia.  The next odd fact is that SN~1998bw had broad [\FeII] lines like
SNe~Ia, but lacked the strong [\FeIII] peak near 4700\AA, which is the strongest
line in the nebular spectra of SNe~Ia. If the [\FeII] identification is correct,
the absence of [\FeIII] clearly must be the result of a different (lower) degree
of ionisation in SN~1998bw. The above facts indicate that something unusual
must be going on with the way the spectrum is powered. 

We used a NLTE code to compute the nebular spectrum of SN~1998bw. The code
computes the deposition of the $\gamma$-rays from the decay of \Cofs\ to \Fefs\ 
in a nebula of uniform density and composition. Together with the positrons, 
which are also produced in the decay of \Cofs\ and which are assumed to be 
trapped, this provides heating to the nebula. 
Cooling by nebular line emission is computed, a matrix of level
populations is solved in NLTE and an emission spectrum is obtained. This 
approximation is reasonable for SN~Ia nebulae (Axelrod 1980), so we apply it
also to SN~1998bw. It is reasonable to do so as long as the nebular spectrum of
SN~1998bw is mostly powered by \Cofs\ decay. Furthermore, the density in the
explosion model CO138, which reproduces the early light curve and spectra of
SN~1998bw, is actually flat below about 12,000~\kms\ (Nakamura \etal 2001a), 
and mixing in the ejecta appears to be required because a stratified abundance
distribution gives nebular lines with a flat-top profile (see Mihalas 1978, 
p.477), which is inconsistent with the observations (Sollerman \etal 2000). 
Mixing was also adopted in the light curve models of IMN98. Masses of
the various elements are given as input to the code.

We modelled 5 spectra of SN~1998bw, dating from three months after maximum,
when the SN was still entering the nebular phase, to more than one year after
maximum. 

Since our code assumes a uniform density, it can be used to determine the outer
velocity of the nebula by fitting the width of the emission features, including
complex blends like [\FeII], which is the broadest feature in all the nebular
spectra of SN~1998bw.  Because line blending is computed, the comparison of 
the widths of synthetic and observed emission features is realistic. 
We found that the velocity of the [\FeII] feature decreased only marginally, 
from 12,000~\kms\ in the first spectrum to 10,000~\kms\ in the last two, 
indicating that $\gamma$-ray deposition slowly became less efficient
in the outer part of the nebula. 

The major difficulties in reproducing the spectra were the presence of the
narrow lines, in particular [\OI] and \MgI], and the apparent absence of
[\FeIII] lines. As for the narrow lines, we chose to fit their peak flux, thus
deriving an upper limit to the masses of these elements.  The absence of
[\FeIII] lines is interesting. Typical models of the nebula have electron
density $n_e \sim 10^7$~cm$^{-3}$, electron temperature $T_e \sim 7000$~K, and
a ratio \FeIII/\FeII\ $\sim 0.5$, so that both [\FeII] and [\FeIII] lines are
strong. This situation is typical also in SNe~Ia. The only way to decrease the
strength of the [\FeIII] lines without also reducing that of the [\FeII] lines 
is of course to decrease the ionisation. This can only be achieved by increasing
$n_e$ to favour recombination. Iron is not the main contributor to the electron
density because of its relatively low abundance, so increasing $n_e$ does not
mean increasing \FeIII/\FeII\ significantly. $n_e$ cannot be increased by
simply increasing the density, as this would imply increasing the mass, which
would result in more emission in all lines. 

One solution is to introduce clumping in the ejecta. This is done using the 
concept of volume filling factor. In practice, the state for a mass $M$ is 
computed using a volume $fV$, where $V$ is the entire volume of the nebula, 
computed from the outer velocity and the time since outburst, and $f$ is the 
volume filling factor, with $0 < f \leq 1$. With this approximation, the 
density of the gas is increased by a factor $f$ to simulate clumping. 
The full volume $V$ is then used to distribute line emissivity and to
compute the spectrum, which corresponds to the assumption of uniformly
distributed clumps. We found that for a filling factor of about 0.1, $n_e$
(averaged over the whole nebula) increases to $\sim 3 \cdot 10^7$~cm$^{-3}$,
while $T_e$ drops to $\sim 6000$~K, producing a lower \FeII/\FeIII\ ratio of
0.05-0.1, which successfully eliminates the [\FeIII] lines, especially at the
more advanced epochs. The presence of clumping is not surprising in the ejecta
of core-collapse SNe (\eg Spyromilio \etal 1993 for SN~1987A and Spyromilio 
1994, Matheson \etal 2000b for SN~1993J). 
Sollerman \etal (2000) using a detailed density
distribution in spherical symmetry obtained a lower degree of ionisation in 
the Fe-rich part of the ejecta, but they had to reduce the ejecta velocity
artificially with respect to the results of the explosion models in
order to obtain a good agreement with the width of the observed lines. 
Our models for the epoch they also modelled, day 139, are similar to theirs, 
providing perhaps a slightly better fit in the blue region where the [\FeIII] 
lines form. We also confirm the absence of lines of \FeI.

The properties of our models fitting the broad [\FeII] feature are summarised in
Table 1, and the synthetic spectra are compared to the observed ones in Figure
1. 

For all the `broad' models, the \Nifs\ mass was about $0.65 \Msun$. This is in
good agreement with the light curve calculations, although it is perhaps a
slight overestimate since there is more flux in the synthetic spectra than
there is in the observed ones, owing to the fact that the narrow observed lines
are reproduced as broad lines. It is also encouraging that the \Nifs\ mass is
very nearly constant over the entire period, showing that the [\FeII] feature is
indeed due to $\gamma$-ray deposition in an expanding nebula.

The O mass, on the other hand, ranges between 3 and $5 \Msun$ in the various
models, increasing first and then decreasing. In the first spectrum the profile
of [\OI] 6300\AA\ is broad, and so our estimate $(M({\rm O}) \sim 3 \Msun)$ is
probably more realistic, considering the large errors which are always incurred
in when determining the mass of O from that line (Schlegel \& Kirshner 1989).
However, at later times the narrow component dominates, our broad synthetic
[\OI] lines have a larger EW than the observed lines, and our O masses are
certainly overestimated. At very late phases the O mass drops again to about 
$3 \Msun$, but this may be just an artifact of using a constant density model 
with a given velocity radius when the real ejecta had a decreasing density
and so effectively a velocity radius decreasing with time.

The fact that the narrow lines of [\OI] and \MgI] decline less rapidly
in luminosity than the [\FeII] feature with time appears to indicate that 
there is a significant amount of these elements at low velocity. 
The ratio of the [\FeII] feature and the broad component of [\OI] 6300\AA\ in 
the spectrum at day 108 also suggests that the Fe/O ratio is larger at high 
velocity. All this is very difficult to reconcile with the classical picture 
of a spherically symmetric explosion, even taking into account mixing.

Sollerman \etal (2000) computed a nebular spectrum for day 139 using the
original density and abundance distributions of model CO138 (IMN98), and 
obtained broad [OI] lines with a flat emission top. This is a consequence of 
distributing O in a shell at high velocity. They overcame this problem by 
mixing the composition to eliminate the flat tops and by arbitrarily 
decreasing the velocity of the outer part of the nebula to reduce the width of 
the [\OI] lines.  However, our code can only handle nebulae of uniform density. 
Thus, in an alternative set of models we tried to fit the narrow-line spectrum, 
assuming it is also powered by $\gamma$-ray deposition. Such models were 
computed only for the last four epochs, since in the first spectrum the narrow 
[\OI] line is still weak. The parameters of the models are detailed in Table 2, 
and the synthetic spectra are shown in Figure 2.

The synthetic narrow-line spectra fit well the lines of \MgI], \NaI~D and [\OI]. 
The width of the lines decreases only slowly with time, as was the case for the
broad [\FeII] feature: velocities range from 7500 to 5000 \kms. The synthetic
[\FeII] feature is in these models significantly too narrow. Therefore we chose
to reproduce its peak only. Once again we found that the \Nifs\ mass is very
nearly constant in all four models, with a value of $\sim 0.35 \Msun$. However,
this is certainly an underestimate, as the total flux in the [\FeII] feature is
not reproduced. Values of the O mass in the low velocity nebula range between
1.5 and $2.7 \Msun$, with the estimate decreasing with time after day +201. O
masses which decrease with time were also derived by Schlegel \& Kirshner
(1989) for the SNe~Ib 1984L and 1985F and for the SN~II 1980K.

An upper limit to the O mass can also be derived from the flux in the [\OI]
6300\AA\ line and the nebular temperature in the high density limit ($n_e \geq
10^6$cm$^{-3}$) using the formula
\begin{equation}
 M({\rm O}) = 10^8 d^2 F([{\rm \OI]}) e^{2.28/T_4}
\end{equation} 
(Uomoto 1986), where $M$(O) is the O mass in $\Msun$, $d$ is the distance to
the  SN in Mpc, F is the flux of the [\OI] line in erg~cm$^{-2}$~s$^{-1}$ and
$T_4$  is the temperature of the nebula in units of $10^4$K.  The high density
limit should apply given the conditions in our model nebulae (Tables 1 and 2).

The values we obtained for SN~1998bw ($d=39.3$Mpc) using values of $T_4$ as
derived from our models are listed in Table 3. The O mass thus derived is in
good agreement with that used in the model calculations, except at the latest
epochs, when the density is lower and the condition of Eq.(1) is not as well
satisfied.

\section{A `portrait' of SN~1998bw} 

Most of the work on SN~1998bw has been based on spherically symmetric models.
The very bright and relatively broad light curve of SN~1998bw can be reproduced
by a family of explosion models with various values of the fundamental
parameters (\Mej, \KE).  All models require $M$(\Nifs)$ \sim 0.4 - 0.7 \Msun$
to power the bright light curve peak. This is about one order of magnitude
larger than in typical core-collapse SNe. Models with different \KE\ yield
different synthetic spectra, and by comparing with the observed early-time
spectra of SN~1998bw and trying to fit the very broad absorption features,
Nakamura \etal (2001a) selected a model with \Mej$ = 10.9 \Msun$, \KE$ = 5 \cdot
10^{52}$~erg (model CO138E50). This model has the same \Mej\ but a larger \KE\ 
than the model CO138 of IMN98, and it yields a better fit to the observed 
bolometric light curve, the evolution of the photospheric velocity and the 
spectra of SN~1998bw. The large value of \KE\ easily qualifies SN~1998bw as 
`the' Type Ic Hypernova. The mass of the exploding CO star was $13.8 \Msun$, 
which implies a main sequence mass of $\sim 40 \Msun$. Similar results were 
also obtained by Branch (2001) from models of the early time spectra.

These explosion models work well around maximum, but the predicted light curve
tail is  quite a bit steeper than the observed one, starting at about 60 days
and continuing until at least day 200. This signals that either these
energetic  models underestimate $\gamma$-ray deposition at those phases or that
an additional source of energy is present, or both.

The nebular spectra of SN~1998bw also had a peculiar evolution. Early on, the
SN showed a `composite' spectrum: [\FeII] lines, typical of SNe~Ia, were strong,
and so were lines of [\OI] and \MgI], which are typical of SNe~Ib/c. At the
same time [\FeIII] lines, also typical of SNe~Ia, were absent. The [\OI] and
\MgI] lines grew stronger with time relative to the [\FeII] lines, but this was
due mostly to a narrow component, which became more and more the dominant one
as time progressed. The emergence of the narrow profiles occured at about the
same time as the light curve deviated from the model prediction.

The broad spectrum could be reproduced with $M$(\Nifs)$ \sim 0.65 \Msun$ and an
outer nebula velocity of $\sim 12,000$\,\kms. This is consistent with the 
distribution of \Nifs\ in CO138E50 of Nakamura \etal (2001a), but the
estimate of the \Nifs\ mass is somewhat larger here. Note, however, that the
light curve obtained with CO138E50 underestimates the late time flux (Nakamura 
\etal 2001a, Fig.5), and so this result is not unexpected. The
narrow lines have a width of only $\sim 6,000$~\kms, and can be reproduced with
a \Nifs\ mass of $0.35 \Msun$. The velocity of the narrow [\OI] line appears to
be consistent with the lower energy model CO138E10 of Nakamura \etal (2001a).

On average, our nebular models required enclosed masses of $\sim 3\Msun$ below
a velocity of $6000$\,\kms\ and of $\sim 5\Msun$ below $12000$\,\kms. Both
these values are larger than the corresponding values from model CO138 by about
$2\Msun$. This may suggest the presence of a high density inner part of the
ejecta, which could explain the slow decline of the light curve between days 
60 and 200, as suggested by Nakamura \etal (2001a) and Chugai (2000). 
A similar suggestion was made by 
Iwamoto \etal (1999) and Mazzali \etal (2000) for the other Type Ic hypernova, 
SN~1997ef. However, the two situations are different.  For SN~1997ef, evidence 
of a high density region at low velocities came from the fact that the 
spectra remain photospheric for a long time, and that the photospheric 
velocity drops below the velocity of the inner cutoff. In a 1-dimensional 
explosion model, this is the separation between the material which is ejected 
by the SN and the infalling matter which forms the compact remnant. 
In SN~1998bw the photospheric phase ends earlier, and the need for the 
central high density region comes from the behaviour of the light curve, 
and it is therefore a less direct result.

Rather strong clumping had to be introduced to increase recombination and keep
the [\FeIII] lines from forming. Introducing clumping may slow the light curve
somewhat. The increasing strength of the low-velocity intermediate-mass element
emission, however, indicates that this will not be sufficient to explain all
observations. The same would apply to the approach adopted by Sollerman \etal
(2000), who reduced the velocity of the ejecta, thus enhancing $\gamma$-ray
deposition, and could thus improve the fit to both the light curve and the
spectrum. Also, with this approach the [\FeII] lines are not expected ever to
become broader than the [\OI] line.

A more complete solution which takes into account all the observed facts should
probably include the presence of asymmetry in the explosion, both geometrical
and in the density distribution. If SN~1998bw was actually linked to GRB980425, 
the fact that the explosion was aymmetric should not be surprising. 
Observations indicate that a rather large mass
of O-dominated material is present at low velocity. Spherical explosions do
not allow that: they are very effective at `emptying' the central region, and
always place the unburned elements at the top of the ejecta. 

A 2-dimensional explosion model was computed by Maeda \etal (2001). 
If the explosion is highly asymmetric, it produces large quantities of unburned
material (mostly C and O) expanding at low velocity in directions away from the 
axis along which most of the energy was released and \Nifs\ synthesised. Our 
vantage point must have been very close to that axis, because we also detected 
the GRB, as Maeda \etal (2001) show from an analysis of the line profiles. 
At early times, the fast-expanding lobes must have been much brighter than the 
rest, and so we observed the broad-lined spectra and the bright light curve. 
The fast-moving regions rapidly became thin, and soon emission lines appeared. 
Initially the emission lines were broad, dominated by the hyper-energetic lobes. 
The distribution of density and abundances in a 2-dimensional model
(Maeda \etal 2001, Fig. 2) shows that in the region of the ejecta with 
velocities of about 10000\,\kms\ the density contours deviate less from
spherical symmetry. Therefore, early in the nebular phase, when these 
fast-moving regions contribute to the nebular spectrum, a sufficient amount of 
high velocity oxygen is present that the [\OI] line can be expected to be broad. 
Later, though, the $\gamma$-rays from the fast-moving \Cofs\ could escape 
that region more and more easily, and the inner, slowly moving, O-dominated 
part of the ejecta would dominate the spectrum. A smaller amount of 
\Cofs\ is present at low velocities, but the $\gamma$-rays it produces are 
trapped more efficiently and excite the O and Mg there. At lower
velocities, though, the distribution of oxygen becomes preferentially
equatorial, and one expects the [\OI] line to decrease significantly
in width as the effective size of the emitting nebula decreases. 
Maeda \etal (2001) suggest that the relative width of the [\FeII] complex 
and the [\OI] line can be explained by such a model.
However, because O dominates over Fe at low velocity, the strength of
the [\OI] line should increase relative to that of the [\FeII]
feature, as is indeed observed. 

The low-velocity cutoff of the asymmetric explosion model (Maeda \etal 2001, 
Fig. 2) is only about 1500\,\kms, which is significantly less than in the 1D 
models (5000\,\kms\ in CO138E50; Nakamura \etal 2001a, Fig.7), also
because of the lower overall \KE\ of the asymmetric models (see below). 
This feature of the 2D models seems to match nicely the need for a
low-velocity, high-density region resulting from the behaviour of the
light curve. 
Additional deposition in the low-velocity region of the $\gamma$-rays produced 
in the high velocity region may in turn increase the deposition function 
above what our spherically symmetric models could estimate, and may help 
explain the slowly declining tail of the light curve.  

That the SN~1998bw explosion was asymmetric is not a new suggestion: first the
polarization measurements indicated an axial ratio of 2-3:1 (H\"oflich \etal
1999), and the calculation of the explosion of a rotating core (McFadyen \&
Woosley 1999) also gave similar results, in an effort to explain the connection
between SN~1998bw and GRB980425. More detailed results have to await detailed
numerical models in two dimensions. If the explosion was asymmetric, our
results for \KE\ in spherical symmetry are overestimated, because those would
only refer to the fast-moving part of the ejecta. In fact, Maeda \etal (2001)
propose a model with \KE $= 10$ foe. However, the estimate of the
\Nifs\ mass in the nebular epoch ($\sim 0.5 \Msun$) is probably correct, as it
should not be significantly influenced by the asymmetry as long as the ejecta 
are completely optically thin. In this case, in fact, every optical photon 
created by the deposition of $\gamma$-rays and positrons escapes the nebula 
immediately and can be observed as SN light. As for the value of \Mej, this 
can only be determined by computing light curves and spectra using 3D 
hydrodynamical models of the explosion (\eg Khokhlov \etal 1999), but a careful 
analysis of the spectra, especially at late times, when both the fast and 
the slow components are observable, can yield at least approximate results.

{\bf Acknowledgements.} 
This work has been supported in part by the Grant-in-Aid for
Scientific Research (12640233) and COE research (07CE2002) of the
Japanese Ministry of Education, Science, Culture, and Sports in Japan.
It is a pleasure to thank Alex Filippenko for enlightening discussions 
on the late-time spectra of SNe~Ic. 
 
%%%%%%%%%%%%  References %%%%%%%%%%%%%%%%%%%%%%%%%%%%%%%%%%%%%%%

%%%%%%%%%%%% Table 1 %%%%%%%%%%%%%%%%%%%%%%%%%%%%%%%%%%%%%%%
\begin{deluxetable}{cccccccccc}
\scriptsize
\tablenum{1}
\tablecaption{Parameters of the `broad' synthetic spectra.}
\tablehead{\colhead{date} &
\colhead{time from max} &
\colhead{SN epoch} &
\colhead{v} &
\colhead{M(\Nifs)} &
\colhead{M(O)} &
\colhead{M(tot)} &
\colhead{D$_{\gamma}$\tablenotemark{\dag}} &
\colhead{T$_e$} &
\colhead{log n$_e$}   \nl
\colhead{~} &
\colhead{days} &
\colhead{days} &
\colhead{km s$^{-1}$} &
\colhead{$\Msun$} &
\colhead{$\Msun$} &
\colhead{$\Msun$} &
\colhead{~~} &
\colhead{K} &    
\colhead{g cm$^{-3}$}  \nl}
\startdata
12 Aug  98 & ~+94 & 108 & 12500 & 0.68 & 3.0 & 5.0 & 0.31 & 6286 & 7.76 \nl 
12 Sept 98 & +125 & 139 & 12000 & 0.62 & 3.4 & 5.2 & 0.24 & 5910 & 7.47 \nl 
26 Nov  98 & +201 & 214 & 11000 & 0.68 & 5.0 & 7.3 & 0.19 & 5092 & 7.06 \nl 
12 Apr  99 & +337 & 349 & 10000 & 0.65 & 3.9 & 5.4 & 0.09 & 4090 & 6.40 \nl 
21 May  99 & +376 & 388 & 10000 & 0.65 & 3.2 & 4.4 & 0.07 & 3874 & 6.22 \nl 
\enddata
\tablenotetext{\dag}{ D$_{\gamma}$ is the $\gamma$-ray deposition fraction}
\end{deluxetable}

%%%%%%%%%%%% Table 2 %%%%%%%%%%%%%%%%%%%%%%%%%%%%%%%%%%%%%%%
\begin{deluxetable}{cccccccccc}
\scriptsize
\tablenum{2}
\tablecaption{Parameters of the `narrow' synthetic spectra.}
\tablehead{\colhead{date} &
\colhead{time from max} &
\colhead{SN epoch} &
\colhead{v} &
\colhead{M(\Nifs)} &
\colhead{M(O)} &
\colhead{M(tot)} &
\colhead{D$_{\gamma}$\tablenotemark{\dag}} &
\colhead{T$_e$} &
\colhead{log n$_e$}   \nl
\colhead{~} &
\colhead{days} &
\colhead{days} &
\colhead{km s$^{-1}$} &
\colhead{$\Msun$} &
\colhead{$\Msun$} &
\colhead{$\Msun$} &
\colhead{~~} &
\colhead{K} &    
\colhead{g cm$^{-3}$}  \nl}
\startdata
12 Sept 98 & +125 & 139 & 7500 & 0.36 & 2.1 & 3.0 & 0.32 & 5971 & 7.76 \nl 
26 Nov  98 & +201 & 214 & 6000 & 0.33 & 2.7 & 3.8 & 0.28 & 5129 & 7.43 \nl 
12 Apr  99 & +337 & 349 & 6000 & 0.37 & 2.0 & 3.2 & 0.12 & 3987 & 6.73 \nl 
21 May  99 & +376 & 388 & 5500 & 0.34 & 1.4 & 2.2 & 0.09 & 3733 & 6.61 \nl 
\enddata
\tablenotetext{\dag}{ D$_{\gamma}$ is the $\gamma$-ray deposition fraction}
\end{deluxetable}

%%%%%%%%%%%% Table 3 %%%%%%%%%%%%%%%%%%%%%%%%%%%%%%%%%%%%%%%
\begin{deluxetable}{cccccc}
\scriptsize
\tablenum{3}
\tablecaption{[\OI] 6300\AA\ fluxes and the O mass.}
\tablehead{\colhead{SN epoch} &
\colhead{model} &
\colhead{F([\OI])} &
\colhead{T$_4$} &
\colhead{M(O)(flux)} &
\colhead{M(O)(model)}    \nl
\colhead{days} & 
\colhead{~} &
\colhead{erg cm$^{-2}$ s$^{-1}$}  &
\colhead{$10^4$K} &
\colhead{$\Msun$} &
\colhead{$\Msun$} \nl}
\startdata
 108 & broad  & 6.0 10$^{-13}$ & 0.63 & 3.2 & 3.0 \nl 
 139 & narrow & 3.3 10$^{-13}$ & 0.60 & 2.1 & 2.1 \nl
 214 & narrow & 1.4 10$^{-13}$ & 0.51 & 1.7 & 2.7 \nl
 349 & narrow & 2.0 10$^{-14}$ & 0.40 & 0.8 & 2.0 \nl
 388 & narrow & 6.0 10$^{-15}$ & 0.37 & 0.4 & 1.4 \nl
\enddata
\end{deluxetable}

%%%%%%%%%%%%%%%%%%%%%%%%%%%%% figures %%%%%%%%%%%%%%%%%%%%%%%%%%%%%%%%%%%

%%%%%% Fig.1
\begin{figure}[t]
\epsfxsize=15.5cm % fix the x-dimension and scales y-dim. to x-dim.
\hspace{3.3cm}
%\epsfbox{sn98bwneballbr.ps}
\epsfbox{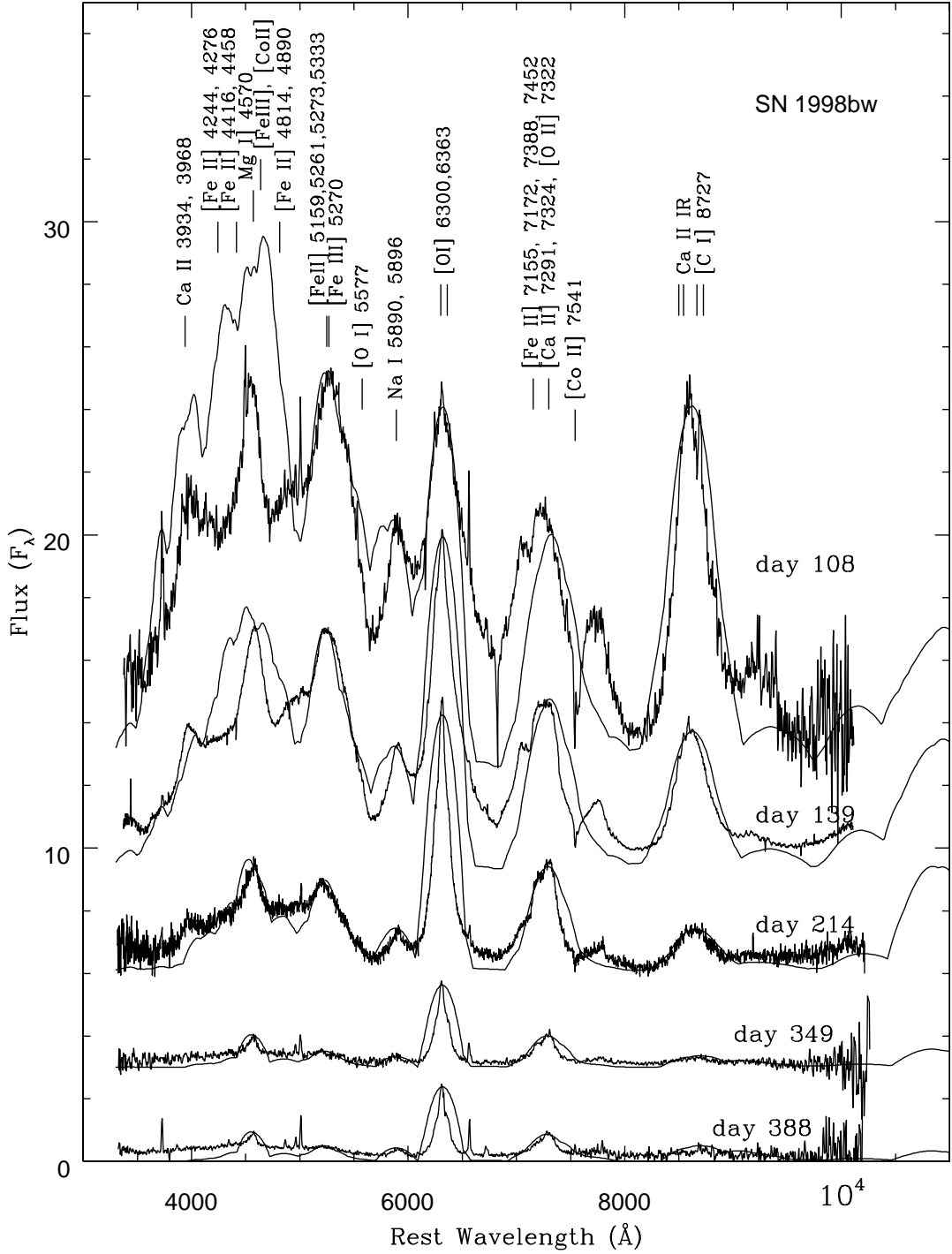}
%for centering: act on hspace argument
\caption[h]{The observed spectra of SN~1998bw from 12 August 1998 to 21 May
1999 compared to the `broad lined' synthetic spectra (Table 1). All spectra
have been shifted arbitrarily.   }
\end{figure}

%%%%%% Fig.2
\begin{figure}[t]
\epsfxsize=15.5cm % fix the x-dimension and scales y-dim. to x-dim.
\hspace{3.3cm}
%\epsfbox{sn98bwneballnar.ps}
\epsfbox{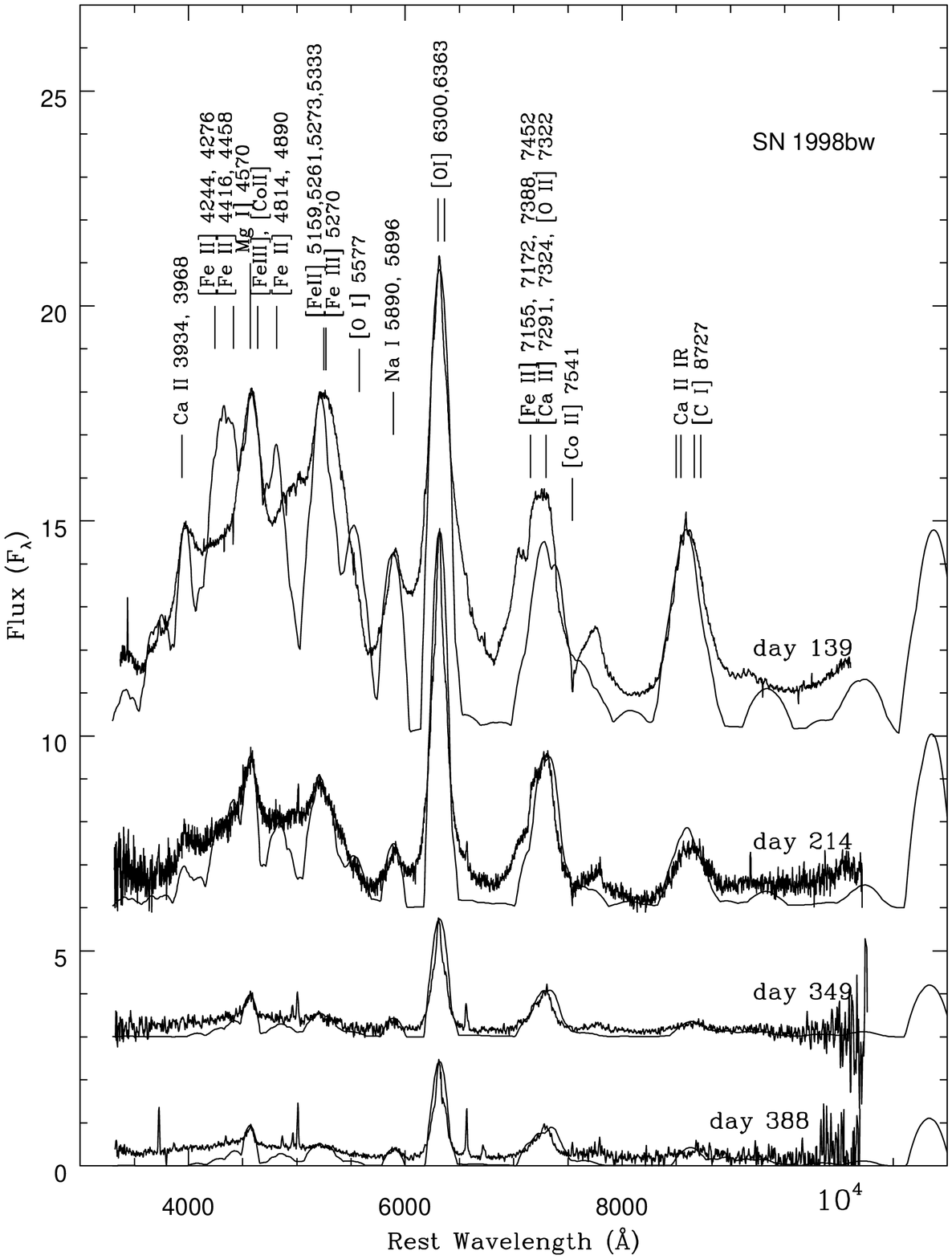}
%for centering: act on hspace argument
\caption[h]{The observed spectra of SN~1998bw from 12 September 1998 to 21 May
1999 compared to the `narrow lined' synthetic spectra (Table 2). All spectra
have been shifted arbitrarily.   }
\end{figure}


\noindent
\begin{thebibliography}{}

\bibitem[Ax80]{ax80} 
Axelrod, T., 1980, Ph.D. Thesis, Univ. of California, Santa Cruz 

\bibitem[Branch]{Branch} 
Branch,~D., 2001, in Supernovae and Gamma Ray Bursts,
ed. M. Livio et al. (Cambridge: Cambridge University Press), in press 

\bibitem[Chu]{Chu} Chugai, N.N., 2000, Astron. Letters 26, 797

\bibitem[Fer]{fer} Ferland, G.J., \& Persson, S.E., 1989, ApJ 347, 656 

\bibitem[Fil]{fil} Filippenko,~A.V., 1989, AJ 97, 726 

\bibitem[FPS]{fps} 
Filippenko,~A.V., Porter, A.C., \& Sargent, W.L.W., 1990, AJ 100, 1575 

\bibitem[F95]{f95} Filippenko,~A.V., et al., 1995, ApJ 450, L11 

\bibitem[Gal98]{gvp98}
Galama, T. J., et al., 1998, Nature 395, 670

\bibitem[Hof99]{hof99} 
H\"{o}flich, P., Wheeler, J.C., \& Wang, L.-F., 1999, ApJ 521, 179

\bibitem[Iwamoto et al.,\ 1998]{imnun98}
Iwamoto, K., et al., 1998, Nature 395, 672

\bibitem[Iwamoto et al.,\ 1999]{imnun99}
Iwamoto, K., et al., 1999, ApJ 534, 660

\bibitem[kay]{kay} Kay, L.E., Halpern, J.P., Leighly, K.M., Heathcote, S.,
Magalhaes,~A.M., \& Filippenko,~A.V., 1998, IAU Circ. 6969 

\bibitem[kh99]{kh99} Khokhlov, A.M., H\"{o}flich, P., Oran, E.S., 
Wheeler, J.C., Wang, L.-F., Chtchelkanova, \& A.Yu., 1999, ApJ 524, L107 

\bibitem[mw99]{mw99}
MacFadyen, A.I. \& Woosley, S.E. 1999, ApJ 524, 262

\bibitem[ms99]{ms99}
McKenzie, E.H., \& Schaefer, B.E. 1999, PASP 111, 964

\bibitem[Mat00a]{mat00a} Matheson, T., Filippenko, A.V., Chornock, R., 
Leonard, D.C., \& Li, W., 2000a, AJ 119, 2303 

\bibitem[Mat00b]{mat00b} Matheson, T., Filippenko, A.V., Ho, L.C., 
Barth, A.J., \& Leonard, D.C., 2000b, AJ 120, 1499 

\bibitem[Mat01]{mat01} Matheson, T., Filippenko, A.V., Leonard, D.C.,
\& Shields, J.C., 2001, AJ 121, 1648 

\bibitem[mae00]{mae00} Maeda, K., Nakamura, T., Nomoto, K., Mazzali, P. A.,
\& Hachisu, I., 2001, ApJ, submitted (astro-ph/0011003)

\bibitem[ma00]{ma00} 
Mazzali, P. A., Iwamoto, K., \& Nomoto, K., 2000, ApJ 545, 407 

\bibitem[mih]{mih} Mihalas, D., 1979, Stellar Atmospheres, 2nd ed.,
Freeman \& Co., San Francisco 

\bibitem[Nak01a]{Nak01a}
Nakamura, T., Mazzali, P.A., Nomoto, K., \& Iwamoto, K., 2001a, ApJ 550, 991
 
\bibitem[Nak01b]{Nak01b} 
Nakamura, T., Umeda, H., Iwamoto, K., Nomoto, K., Hashimoto, M., 
Hix, R.W., Thielemann, F.-K., 2001b, ApJ 555, in press

\bibitem[Nom94]{nom94} 
Nomoto, K., Yamaoka, H., Pols, O.R., van den Heuvel, E.P.J., Iwamoto, K., 
Kumagai, S., \& Shigeyama, T., 1994, Nature 371, 227

\bibitem[Nom01]{nom01} 
Nomoto, K., \etal, 2001, in Supernovae and Gamma Ray Bursts,
ed. M. Livio et al. (Cambridge: Cambridge University Press), in press 
(astro-ph/0003077)

\bibitem[Patat et al.,\ 2001] {p01}
Patat, F. et al., 2001, ApJ 555, in press, (astro-ph/0103111)

\bibitem[rich]{rich} 
Richmond, M.W., \etal, 1996, AJ 111, 327 

\bibitem[sk89]{sk89}
Schlegel, E.M., \& Kirshner, R.P., 1989, AJ 98, 577

\bibitem[s00]{s00}
Sollerman, J., Kozma, C., Fransson, K., Leibundgut, B., Lundqvist, P., Ryde, F.,
\& Woudt, P., 2000, ApJ 537, L127

\bibitem[spy93]{spy93} 
Spyromilio, J., Stathakis, R.A., \& Meurer, G.R., 1993, MNRAS 263, 530 

\bibitem[spy94]{spy94} Spyromilio, J., 1994, MNRAS 266, 61 

\bibitem[uo]{uo} Uomoto, A., 1986, ApJ 310, L35

\end{thebibliography}
\end{document}